\documentclass[prl,twocolumn,showpacs,preprintnumbers,amsmath,amssymb]{revtex4}
\usepackage{graphicx}
\usepackage{dcolumn}
\usepackage{bm}
\begin{document}
\preprint{PRL/J\"ulich-Ferrara-Erlangen}
\title{A Method to Polarize Stored Antiprotons to a High Degree}
\author{F. Rathmann$^1$}\email{f.rathmann@fz-juelich.de}
\author{P. Lenisa$^2$}
\author{E. Steffens$^3$}
\author{M. Contalbrigo$^2$}
\author{P.F. Dalpiaz$^2$}
\author{A. Kacharava$^3$}
\author{ A. Lehrach$^1$}
\author{B. Lorentz$^1$}
\author{R. Maier$^1$}
\author{D. Prasuhn$^1$}
\author{H. Str\"oher$^1$}
\affiliation{$^1$ Institut f\"ur Kernphysik, Forschungzentrum J\"ulich, 52428 J\"ulich, Germany}
\affiliation{$^2$Universit\`a di Ferrara and INFN, 44100 Ferrara, Italy}
\affiliation{$^3$Physikalisches Institut II, Universit\"at Erlangen--N\"urnberg, 91058 Erlangen, Germany}

\date{\today}
\begin{abstract}
  Polarized antiprotons can be produced in a storage ring by
  spin--dependent interaction in a purely electron--polarized hydrogen
  gas target.  The polarizing process is based on spin transfer from
  the polarized electrons of the target atoms to the orbiting
  antiprotons. After spin filtering for about two beam lifetimes at
  energies $T\approx 40-170$~MeV using a dedicated large acceptance
  ring, the antiproton beam polarization would reach $P=0.2-0.4$.
  Polarized antiprotons would open new and unique research opportunities for
  spin--physics experiments in $\bar{p}p$ interactions.

\end{abstract}
\maketitle


For more than two decades, physicists have tried to produce beams of
polarized antiprotons.  Conventional methods like atomic beam sources
(ABS), appropriate for the production of polarized protons and heavy
ions cannot be applied, since antiprotons annihilate with matter.
Polarized antiprotons have been produced from the decay in flight of
$\bar{\Lambda}$ hyperons at Fermilab. The achieved intensities with
antiproton polarizations $P>0.35$ never exceeded $1.5 \cdot
10^5$~s$^{-1}$ \cite{grosnick}.  Scattering of antiprotons off a
liquid hydrogen target could yield polarizations of $P\approx 0.2$,
with beam intensities of up to $2 \cdot 10^3$~s$^{-1}$ \cite{spinka}.
Unfortunately, both approaches do not allow efficient accumulation in
a storage ring, which would greatly enhance the luminosity.  Spin
splitting using the Stern--Gerlach separation of the given magnetic
substates in a stored antiproton beam was proposed in 1985
\cite{niinikoski}.  Although the theoretical understanding has much
improved since then \cite{cameron}, spin splitting using a stored beam
has yet to be observed experimentally.

Interest in the polarization of antiprotons has recently been
stimulated by a proposal to build a High Energy Storage Ring (HESR)
for antiprotons at the new Facility for Antiproton and Ion Research
(FAIR) at the Gesellschaft f\"ur Schwer\-ionenforschung (GSI) in
Darmstadt \cite{cdr-gsi}.  A Letter--of--Intent for spin--physics
experiments has been submitted by the PAX collaboration \cite{pax-loi}
to employ a polarized antiproton beam incident on a polarized internal
storage cell target \cite{steffens}.  A beam of polarized antiprotons
would enable new experiments, such as the first direct measurement of the
transversity distribution of the valence quarks in the proton, a test
of the predicted opposite sign of the Sivers--function --- related to the
quark distribution inside a transversely polarized nucleon --- in
Drell--Yan as compared to semi--inclusive deep--inelastic scattering,
and a first measurement of the moduli and the relative phase of the
time--like electric and magnetic form factors $G_{\mathrm{E,M}}$ of
the proton \cite{pax-loi}.  

In 1992 an experiment at the Test Storage Ring (TSR) at MPI Heidelberg
showed that an initially unpolarized stored 23~MeV proton beam can be
polarized by spin--dependent interaction with a polarized hydrogen gas
target \cite{rathmann,zapfe,zapfe2}.  In the presence of polarized
protons of magnetic quantum number ${m=\frac{1}{2}}$ in the target,
beam protons with ${m=\frac{1}{2}}$ are scattered less often, than
those with ${m=-\frac{1}{2}}$, which eventually caused the stored beam
to acquire a polarization parallel to the proton spin of the hydrogen
atoms during spin filtering.  
In an analysis by Meyer three different mechanisms were identified,
that add up to the measured result \cite{meyer}.  One of these
mechanisms is spin transfer from the polarized electrons of the
hydrogen gas target to the circulating protons.  Horowitz and Meyer
derived the spin transfer cross section ${p+\vec{e}\rightarrow
  \vec{p}+e}$ (using ${c=\hbar=1}$) \cite{horowitz-meyer},
\begin{equation}
        \sigma_{e_{||}} 
= -\frac{4 \pi \alpha^2 (1+a) 
                          m_e  
                          }{p^2  m_p}\cdot C_0^2 \cdot
                          \frac{v}{2 \alpha}
                          \cdot
                          \sin\left( \frac{2 \alpha}{v} \ln(2pa_0)\right)\;,
        \label{eq:sigma_long}
\end{equation}
where ${\alpha}$ is the fine--structure constant, $a$
is the anomalous magnetic moment of the proton, ${m_e}$ and
${m_p}$ are the rest mass of electron and proton, ${p}$ is the
momentum in the CM system, ${a_0=52900 \;{\rm fm}}$ is the Bohr radius
and ${C_0^2=2\pi\eta/[\exp(2\pi\eta)-1]}$ is the square of the Coulomb
wave function at the origin.  The Coulomb parameter ${\eta}$ is given
by ${\eta = -z\alpha/v}$ (for antiprotons, $\eta$ is positive). ${z}$
is the beam charge number and ${v}$ the relative velocity of particle
and projectile in the laboratory system.

\begin{figure}[htb]
\begin{center}
  \includegraphics[width=7.0cm]{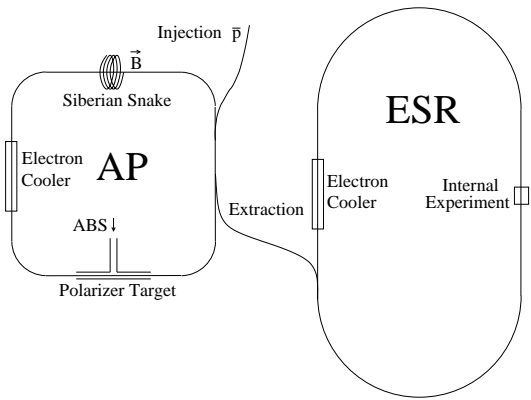}
\caption{\label{fig:sketch}Antiproton polarizer (AP) and 
  experimental storage ring (ESR).}
\end{center}
\end{figure}

In the following we evaluate a concept for a dedicated antiproton
polarizer ring (AP).  Antiprotons would be polarized by the
spin--dependent interaction in an electron--polarized hydrogen gas
target. This spin--transfer process is {\it calculable}, whereas, due
to the absence of polarized antiproton beams in the past, a
measurement of the spin--dependent $\bar{p}p$ interaction is still
lacking, and only theoretical models exist \cite{mull}.  The polarized
antiprotons would be subsequently transferred to an experimental
storage ring (ESR) for measurements (Fig.~\ref{fig:sketch}).
Both the AP and the ESR should be operated as synchrotrons with
electron cooling to counteract emittance growth.  In both rings the
beam polarization should be preserved during acceleration without loss
\cite{pol-conservation}.  The longitudinal spin--transfer cross
section is twice as large as the transverse one \cite{meyer},
$
\sigma_{e_{\parallel}}=2\cdot \sigma_{e_{\bot}}
$, the stable spin direction of the beam at the location of the
polarizing target should therefore be longitudinal as well, which
requires a Siberian snake in a straight section opposite the
polarizing target \cite{snakes}.

A hydrogen gas target of suitable substate population represents a
dense target of quasi--free electrons of high polarization and areal
density. Such a target can be produced by injection of two hyperfine
states with magnetic quantum numbers $|m_J=+\frac{1}{2},
m_I=+\frac{1}{2}\rangle$ and $|+\frac{1}{2}, -\frac{1}{2}\rangle$ into
a strong longitudinal magnetic holding field of about $B_{||}=300$~mT.
The maximum electron and nuclear target polarizations in such a field
are $Q_e=
0.993$ and $Q_z=
0.007$ \cite{haeberli}.
Polarized atomic beam sources presently produce a flux of hydrogen
atoms of about $q=1.2\cdot10^{17}$~atoms/s in two hyperfine states
\cite{zelenski}.  Our model calculation for the polarization buildup
assumes a moderate improvement of 20\%, i.e.  a flow of
$q=1.5\cdot10^{17}$~atoms/s.

The beam lifetime in the AP can be expressed as function of the
Coulomb--Loss cross section $\Delta \sigma_C$ and the total hadronic
$\bar{p}p$ cross section $\sigma_{\mathrm{tot}}$,
\begin{eqnarray}
\tau_{\mathrm{AP}}
 =\frac{1}{(\Delta \sigma_C + 
 \sigma_{\mathrm{tot}}) \cdot d_t 
 \cdot f_{\mathrm{AP}}}\;.
\label{eq:tau_ap}
\end{eqnarray}

The density $d_t$ of a storage cell target depends on the flow of
atoms $q$ into the feeding tube of the cell, its length along the beam
$L_{\mathrm{beam}}$, and the total conductance $C_{\mathrm{tot}}$ of
the storage cell 
$
d_t=\frac{1}{2}\frac{L_{\mathrm{beam}}\cdot q}{C_{\mathrm{tot}}}
$ \cite{steffens}. The conductance of a cylindrical tube
$C_{\mathrm{\circ}}$ for a gas of mass $M$ in the regime of molecular
flow (mean free path large compared to the dimensions of the tube) as
function of its length $L$, diameter $d$, and temperature $T$, is
given by $
C_{\mathrm{\circ}}=3.8\cdot \sqrt{\frac{T}{M}}\cdot
\frac{d^3}{L+\frac{4}{3}\cdot d}.
\label{eq:C_circ}
$ The total conductance $C_{\mathrm{tot}}$ of the storage cell is
given by $
C_{\mathrm{tot}} = C_{\mathrm{\circ}}^{\mathrm{feed}} + 2\cdot
C_{\mathrm{\circ}}^{\mathrm{beam}},
$ where $C_{\mathrm{\circ}}^{\mathrm{feed}}$ denotes the conductance
of the feeding tube and $C_{\mathrm{\circ}}^{\mathrm{beam}}$ the
conductance of one half of the beam tube.  The diameter of the beam
tube of the storage cell should match the ring acceptance angle
$\Psi_{\mathrm{acc}}$ at the target, $
d_{\mathrm{beam}}=2\cdot \Psi_{\mathrm{acc}}\cdot \beta
$, where for the $\beta$--function at the target, we use
$\beta=\frac{1}{2}\,L_{\mathrm{beam}}$.  One can express the target
density in terms of the ring acceptance, $d_t \equiv
d_t(\Psi_{\mathrm{acc}})$, where the other parameters used in the
calculation are listed in Table~\ref{tab:polarizer_section}.
\begin{table}
\begin{tabular}{l|l|l}
  \hline\hline
  circumference of AP         & $L_{\mathrm{AP}}$      & 150 m  \\
  $\beta$--function at target  & $\beta$                & 0.2 m  \\
  radius of vacuum chamber    & $r$                    & 5 cm   \\
  gap height of magnets       & $2\, g$                & 14 cm \\\hline
  ABS flow into feeding tube  & $q$                    & $1.5 \cdot 10^{17}$ atoms/s \\ 
  storage cell length         & $L_{\mathrm{beam}}$    & 40 cm  \\
  feeding tube diameter       & $d_{\mathrm{feed}}$    & 1 cm   \\
  feeding tube length         & $L_{\mathrm{feed}}$    & 15 cm  \\
  longitudinal holding field  & $B_{||}$               & 300 mT \\
  electron polarization       & $Q_e$                  & 0.9    \\    
  cell temperature            & $T$                    & 100 K  \\\hline\hline 
\end{tabular}
\caption{\label{tab:polarizer_section}Parameters of the AP and the polarizing target section. }
\end{table}

\begin{figure}[htb]
\begin{center}
  \includegraphics[width=7.0cm]{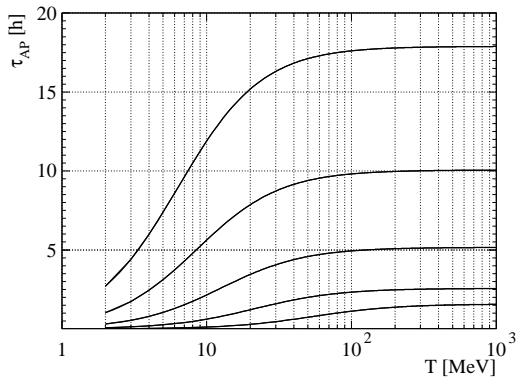}
\caption{\label{fig:tau_ap}Beam lifetime in the AP as function of 
  kinetic energy $T$.  From top to bottom the lines denote
  $\Psi_{\mathrm{acc}}=50$, 40, 30, 20, and 10~mrad.}
\end{center}
\end{figure}

The Coulomb--loss cross section $\Delta \sigma_C$ (using
${c=\hbar=1}$) can be derived analytically in terms of the square of
the total energy $s$ by integration of the Rutherford cross section,
taking into account that only those particles are lost that undergo
scattering at angles larger than $\Psi_{\mathrm{acc}}$,
\begin{eqnarray}
\Delta \sigma_C(\Psi_{\mathrm{acc}})=4\pi\alpha^2 
\frac{(s-2m_{\bar{p}}^2)^2 \, 4m_{\bar{p}}^2}
{s^2(s-4m_{\bar{p}}^2)^2}
\,
\left(\frac{1}{\Psi_{\mathrm{acc}}^2} - \frac{s}{4m_{\bar{p}}^2}\right)\;.
\end{eqnarray}
The total hadronic cross section is parameterized using a function
inversely proportional to the Lorentz parameter
$\beta_{\mathrm{lab}}$.  Based on the $\bar{p}p$ data \cite{PDG} the
parameterization
\begin{eqnarray}
\sigma_{\mathrm{tot}}=\frac{75.5}{\beta_{\mathrm{lab}}}\;(\mathrm{mb})
\end{eqnarray}
yields a description of $\sigma_{\mathrm{tot}}$ with $\approx 15$\%
accuracy up to $T\approx 1000$~MeV.  The AP revolution frequency is
given by
\begin{eqnarray}  
f_{\mathrm{AP}}=\frac{\beta_{\mathrm{lab}} \cdot c}{L_{\mathrm{AP}}}\;.
\label{eq:f_ap}
\end{eqnarray}
The resulting beam lifetime in the AP as function of the kinetic
energy $T$ is depicted in Fig.~\ref{fig:tau_ap} for different
acceptance angles $\Psi_{\mathrm{acc}}$.

The buildup of polarization due to the spin--dependent $\bar{p} e$
interaction in the target [Eq.~(\ref{eq:sigma_long})] as function of
time $t$ is described by
\begin{eqnarray}
P(t)=\tanh\left(\frac{t}{\tau_p}\right)\;,\, \mathrm{where} \;
\tau_p=\frac{1}{\sigma_{e_{\parallel}} \, d_t \, f_{\mathrm{AP}} \, Q_e}
\label{eq:p_of_t}
\end{eqnarray}
denotes the polarization buildup time. The time dependence of the beam
intensity is described by
\begin{eqnarray}
I(t)=I_0 \cdot \exp{\left(-\frac{t}{\tau_{\mathrm{AP}}}\right)} 
         \cdot \cosh{\left(\frac{t}{\tau_p}\right)}\;,
\label{eq:I_of_t}
\end{eqnarray}
where $I_0=N_{\bar{p}}^{\mathrm{AP}}\cdot f_{\mathrm{AP}}$.

\begin{figure}[htb]
\begin{center}
  \includegraphics[width=7.0cm]{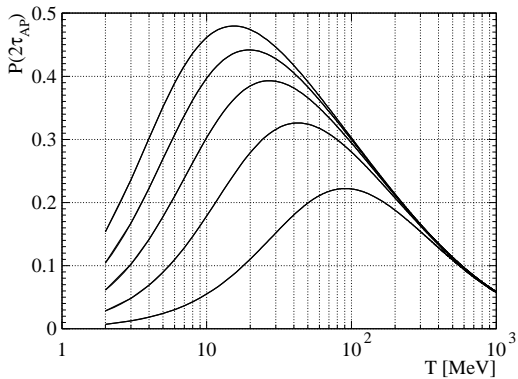}
\caption{\label{fig:pol}Antiproton beam polarization $P(2 \cdot 
  \tau_{\mathrm{AP}})$ [Eq.~(\ref{eq:p_of_t})] as function of beam
  energy for different acceptance angles $\Psi_{\mathrm{acc}}$.
  (Lines are organized as in Fig.~\ref{fig:tau_ap}.)}
\end{center}
\end{figure}

The quality of the polarized antiproton beam can be expressed in terms
of the {\it Figure of Merit} \cite{ohlsen}
\begin{eqnarray}
\mathrm{FOM}(t)=P(t)^2 \cdot I(t)\;.
\label{eq:fom}
\end{eqnarray}
The optimum interaction time $t_{\mathrm{opt}}$, where
$\mathrm{FOM}(t)$ reaches the maximum, is given by $\frac{\mathrm{d}}
{\mathrm{d}\,t} \mathrm{FOM}(t) =0$.  For the situation discussed
here, $t_{\mathrm{opt}}=2 \cdot \tau_{\mathrm{AP}}$ constitutes a good
approximation that deviates from the true values by at most 3\%. The
magnitude of the antiproton beam polarization $P(t_{\mathrm{opt}})$
based on electron spin transfer [Eq.~(\ref{eq:p_of_t})] is depicted in
Fig.~\ref{fig:pol} as function of beam energy $T$ for different
acceptance angles $\Psi_{\mathrm{acc}}$.

The number of antiprotons stored in the AP may be limited by
space--charge effects. With an antiproton production rate of
$R=10^7$~$\bar{p}/s$, the number of antiprotons available at the
beginning of the filtering procedure corresponds to
\begin{eqnarray}
N_{\bar{p}}^{\mathrm{AP}}(t=0)=R \cdot 2 \cdot \tau_{\mathrm{AP}}\;.
\label{eq:n_ap}
\end{eqnarray}
The individual particle limit in the AP is given by \cite{bovet}
\begin{eqnarray}
N_{\mathrm{ind.}}=2\, \pi \, \varepsilon \, \beta_{\mathrm{lab}}^2 \,
\gamma_{\mathrm{lab}}^3 \, (r_p \, F)^{-1} \, \Delta Q\;,
\label{eq:n_ind}
\end{eqnarray}
where $\varepsilon=\Psi_{\mathrm{acc}}^2\cdot \beta$ denotes the
vertical and horizontal beam emittance, $\beta_{\mathrm{lab}}$ and
$\gamma_{\mathrm{lab}}$ are the Lorentz parameters, $r_p=1.5347 \cdot
10^{-18}$~m is the classical proton radius, and $\Delta Q=0.01$ is the
allowed incoherent tune spread. The form factor $F$ for a circular vacuum
chamber \cite{bovet} is given by $F=1+\left(a_y \cdot \frac{a_x +
    a_y}{r^2}\right) \cdot \varepsilon_2 \cdot
(\gamma_{\mathrm{lab}}^2-1) \cdot \frac{r^2}{g^2}$, where the mean
semi--minor horizontal $(x)$ and vertical $(y)$ beam axes
$a_{x,y}=\sqrt{\varepsilon \cdot \beta_{x,y}}$ are calculated from the
mean horizontal and vertical $\beta$--functions
$\beta_{x,y}=L_{\mathrm{AP}}\cdot (2\pi\nu)^{-1}$ for a betatron--tune
$\nu=3.6$.  For a circular vacuum chamber and straight magnet pole pieces
the image force coefficient $\varepsilon_2=0.411$.  The parameter $r$
denotes the radius of the vacuum chamber and $g$ half of the height of
the magnet gaps (Table~\ref{tab:polarizer_section}).

\begin{figure}[htb]
  \includegraphics[width=7.0cm]{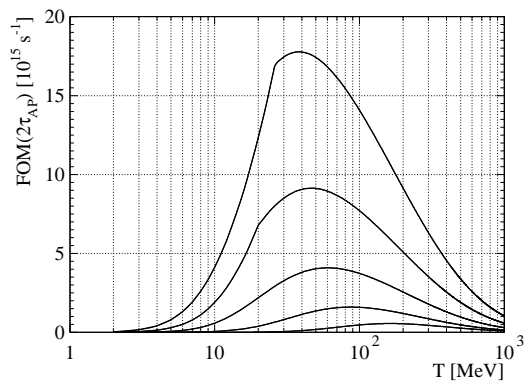}
\caption{\label{fig:fom} Figure of Merit for the polarized antiproton beam for 
  filtering times $t=2\cdot \tau_{\mathrm{AP}}$ as function of beam
  energy. The parameters associated with the maxima are summarized in
  Table~\ref{tab:opt_energies}. (Lines are organized as in
  Fig.~\ref{fig:tau_ap}.)}
\end{figure}

The optimum beam energies for different acceptance angles at which the
polarization buildup works best, however, cannot be obtained from the
maxima in Fig.~\ref{fig:pol}. In order to find these energies, one has
to evaluate at which beam energies the FOM [Eq.~(\ref{eq:fom})],
depicted in Fig.~\ref{fig:fom}, reaches a maximum. The optimum beam
energies for polarization buildup in the AP are listed in
Table~\ref{tab:opt_energies}. The limitations due to space--charge,
$N_{\bar{p}}^{\mathrm{AP}}>N_{\mathrm{ind.}}$ [Eqs.~(\ref{eq:n_ap},
\ref{eq:n_ind})], are visible as kinks in Fig.~\ref{fig:fom} for the
acceptance angles $\Psi_{\mathrm{acc}}=40$ and 50~mrad, however, the
optimum energies are not affected by space--charge.
\begin{table}
\begin{tabular}{c|c|c|c}
$\Psi_{\mathrm{acc}}$ (mrad) & $T$ (MeV) & $\tau_{\mathrm{AP}}$ (h) & $P(2\,\tau_{\mathrm{AP}})$ \\ \hline\hline
10                           & 167       & 1.2                       & 0.19 \\
20                           & 88        & 2.2                       & 0.29 \\
30                           & 61        & 4.6                       & 0.35 \\
40                           & 47        & 9.2                       & 0.39 \\
50                           & 39        & 16.7                      & 0.42 \\ \hline\hline
\end{tabular}
\caption{\label{tab:opt_energies}Kinetic beam energies where the 
 polarized antiproton beam  in the AP reaches the maximum FOM 
for different acceptance angles. }
\end{table}

Spin filtering in a {\it pure} electron target greatly reduces the beam
losses, because $\sigma_{\mathrm{tot}}$ disappears and Coulomb
scattering angles in $\bar{p}e$ collisions do not exceed
$\Psi_{\mathrm{acc}}$ of any storage ring.  With stationary electrons
stored in a Penning trap, densities of about $10^{12}$
electrons/cm$^2$ may be reached in the future \cite{traps}. A typical
electron cooler operated at 10~kV with polarized electrons of
intensity $\approx 1$~mA ($I_e \approx 6.2\cdot 10^{15}$~electrons/s)
\cite{pol-el-sources}, $A=1$~cm$^2$ cross section, and $l=5$~m length
reaches $d_t=I_e \cdot l \cdot (\beta_{\mathrm{lab}}\, c\,A)^{-1}= 5.2
\cdot 10^{8}$~electrons/cm$^2$, which is six orders of magnitude short
of the electron densities achievable with a neutral hydrogen gas
target.  For a pure electron target the spin transfer cross section
is $\sigma_{e_{||}}=670$~mb (at $T = 6.2$~MeV) \cite{horowitz-meyer},
about a factor $15$ larger than the cross sections associated
to the optimum energies using a gas target
(Table~\ref{tab:opt_energies}). One can therefore conclude that with
present day technologies, both above discussed alternatives are no
match for spin filtering using a polarized gas target.

In order to estimate the luminosities available for the ESR, we use
the parameters of the HESR ($L_{\mathrm{HESR}}=440$~m).  After spin
filtering in the AP for $t_{\mathrm{opt}}=2 \cdot \tau_{\mathrm{AP}}$,
the number of polarized antiprotons transfered to HESR is
$N_{\bar{p}}^{\mathrm{AP}}(t=0)/e^2$ [Eq.~(\ref{eq:n_ap})].  The
beam lifetime in the HESR at $T=15$~GeV for an internal polarized
hydrogen gas target of $d_t=7 \cdot 10^{14}$~cm$^{-2}$ is about
$\tau_{\mathrm{HESR}} = 12$~h [Eqs.~(\ref{eq:tau_ap}, \ref{eq:f_ap})],
where the target parameters from Table~\ref{tab:polarizer_section}
were used, a cell diameter $d_{\mathrm{beam}}=0.8$~cm, and
$\sigma_{\mathrm{tot}}= 50$~mb.  Subsequent transfers from the AP to
the HESR can be employed to accumulate antiprotons.  Eventually, since
$\tau_{\mathrm{HESR}}$ is finite, the average number of antiprotons
reaches equilibrium, $
\overline{N_{\bar{p}}^{\mathrm{HESR}}}=R/e^2 \cdot
\tau_{\mathrm{HESR}}= 5.6\cdot 10^{10}
$, independent of $\tau_{\mathrm{AP}}$. An average luminosity of $
\bar{{\cal L}}=
R/(e^2 \cdot \sigma_{\mathrm{tot}})= 2.7 \cdot 10^{31} \;
\mathrm{cm}^{-2}\mathrm{s}^{-1}\
$ can be achieved, with antiproton beam polarizations depending on the
AP acceptance angle $\Psi_{\mathrm{acc}}$
(Table~\ref{tab:opt_energies}).


We have shown that with a dedicated large acceptance antiproton
polarizer ring ($\Psi_{\mathrm{acc}}=10$ to 50~mrad), beam
polarizations of $P=0.2$ to 0.4 could be reached. The energies at
which the polarization buildup works best range from $T=40$ to
170~MeV.  In equilibrium, the average luminosity for
double--polarization experiments in an experimental storage ring (e.g.
HESR) after subsequent transfers from the AP could reach $\bar{{\cal L}}=2.7
\cdot 10^{31} \; \mathrm{cm}^{-2}\mathrm{s}^{-1}$.

The antiproton polarizer, discussed here, would provide highly
polarized antiproton beams of unprecedented quality. In particular the
implementation of this option at the Facility for Antiproton and Ion
Research would open new and unique research opportunities for
spin--physics experiments in $\bar{p}p$ interactions at the HESR.

We would like to thank J. Haidenbauer and N.N. Nikolaev for many
insightful discussions on the subject.



\begin{thebibliography}{99}
\bibitem{pol-conservation}Preservation of beam polarization is
  routinely accomplished at COSY--J\"ulich \cite{lehrach}, and at
  higher energies at the AGS of BNL \cite{ags-bnl}.
\end{thebibliography}
\end{document}